\documentclass[12pt]{article}

\usepackage{authblk}
\usepackage{amsmath,amssymb}
\usepackage{graphicx}
\usepackage{lineno}
\usepackage[margin=1in]{geometry}
\usepackage{hyperref}
\usepackage{booktabs}

\newcommand{\Lex}{\mathcal{L}_{\mathrm{ex}}}
\newcommand{\Qex}{\mathcal{Q}_{\mathrm{ex}}}
\newcommand{\NSex}{\mathcal{NS}_{\mathrm{ex}}}
\newcommand{\PQex}{\mathcal{PQ}_{\mathrm{ex}}}

\newcommand{\Qoneex}{\mathcal{Q}_{1,\mathrm{ex}}}
\newcommand{\PQcert}{\mathcal{PQ}_{1,\mathrm{ex}}}

\newcommand{\Pyr}{\mathrm{Pyr}}

\newcommand{\Qpairex}{\mathcal{Q}_{\mathrm{pair},\mathrm{ex}}}
\newcommand{\PQpair}{\mathcal{PQ}_{\mathrm{pair},\mathrm{ex}}}
\newcommand{\CHSH}{\mathrm{CHSH}}

\title{
Detectability of Post-Quantum Correlations in Symmetry-Restricted Bell Scenarios
}

\author[1]{Marek Gazdzicki}
\author[1]{Francesco Giacosa}
\affil[1]{Jan Kochanowski University, Kielce, Poland}

\date{}

\begin{document}
\maketitle



\begin{abstract}
Searches for post-quantum correlations depend on two distinct ingredients:
how abundant such correlations are in the underlying correlation space and
how efficiently the chosen representation preserves their distinguishability
from quantum correlations. We quantify these effects by introducing the
popularity rate and the projection acceptance, whose product defines the
search performance. We apply these concepts in binary-outcome Bell
scenarios with symmetry-related sites, using the CHSH and pyramid
representations. Particular attention is paid to the distinction between
exchange-symmetric synchronous correlations and correlations satisfying only
exchange symmetry. In the synchronous quantum sector, the
common-vector representation follows automatically from synchrony, whereas
exchange symmetry alone defines a substantially
larger correlation space. We show that this distinction can qualitatively
change the effectiveness of a projected search: the pyramid representation,
which retains substantial discriminatory power in the synchronous sector, has
no discriminatory power when only exchange symmetry is imposed. Our results
provide a simple framework for assessing and comparing
searches for post-quantum correlations.

\end{abstract}

\newpage


\section{Introduction}
\label{sec:introduction}

Bell's theorem~\cite{Bell:1964kc,Bell:1975uz} establishes that correlations
predicted by quantum mechanics cannot, in general, be reproduced by local
hidden-variable theories. In a bipartite Bell experiment, two observers
independently select measurement settings and record the joint distribution of
their outcomes. Depending on the physical assumptions, the resulting
conditional distributions belong to the local, quantum, or no-signalling sets,
denoted by $\mathcal{L}$, $\mathcal{Q}$, and $\mathcal{NS}$,
respectively~\mbox{\cite{Werner:2001cgs,Navascues:2007zcm,Masanes:2005njk}}. They obey
$\mathcal{L}\subseteq\mathcal{Q}\subseteq\mathcal{NS}$. The post-quantum region
$\mathcal{PQ}=\mathcal{NS}\setminus\mathcal{Q}$ therefore contains
no-signalling correlations that cannot be reproduced by quantum
theory~\cite{Popescu:1994kjy}.

The geometry of these correlation sets has been studied extensively in a wide
range of Bell setups~\cite{Brunner:2013est}. Their dimensions, convexity,
facets, and extremal points describe the structure of the complete correlation
space. Experimental analyses, however, rarely retain this complete
information. Instead, the measured probabilities are usually mapped to a
lower-dimensional representation. The best-known example is the
Clauser--Horne--Shimony--Holt (CHSH) correlator~\cite{CHSH1969}, which maps the
scenario with two settings and two outcomes, $(2,2,2,2)$, to one real coordinate. Another example is the
three-dimensional pyramid representation of the scenario with
three settings and two outcomes, $(3,3,2,2)$, recently introduced in Ref.~\cite{Gazdzicki:2026grg}.

Such projections simplify theoretical analyses and experimental tests, but
dimensional reduction necessarily discards information. A post-quantum point
may be mapped into the projected quantum region and thereby become
indistinguishable from a quantum correlation. The effectiveness of a search
therefore depends both on how abundant post-quantum correlations are in the
complete reference space and on what fraction remain distinguishable after
projection.

Previous work has investigated relative volumes of correlation sets and the
typicality or detectability of non-local correlations in different Bell
scenarios~\mbox{\cite{Camillo:2023hkz,Barasinski:2021vzp,deRosier:2019tud,Lin:2021bne}}.
Here we introduce three related quantities. The popularity rate measures the
abundance of post-quantum correlations in the chosen reference space, the
projection acceptance quantifies the fraction that remains identifiable after
projection, and their product defines the search performance. We use
these concepts for the $(2,2,2,2)$ and $(3,3,2,2)$ scenarios by comparing
exchange-symmetric synchronous correlations with correlations satisfying
exchange symmetry alone.

The paper is organised as follows. Section~\ref{sec:geometry} introduces the relevant Bell
correlation spaces and their projections. Section~\ref{sec:popularity} defines the popularity
rate, projection acceptance, and search performance. Section~\ref{sec:results} presents results on the detectability of post-quantum correlations. Section~\ref{sec:conclusions} summarises the paper and draws conclusions.

\section{Bell Correlation Spaces and Projections}
\label{sec:geometry}

\subsection{Correlation spaces}

Consider a bipartite Bell scenario in which each party chooses one of finitely
many measurement settings and obtains one of finitely many outcomes. The
experimentally accessible information is contained in the conditional
probability distribution
\[
P(a,b|x,y)~,
\]
where $x$ and $y$ denote the measurement settings and $a$ and $b$ the corresponding outcomes.

The complete probability space is constrained by positivity,
\[
P(a,b|x,y)\ge0~,
\]
and normalisation,
\[
\sum_{a,b}P(a,b|x,y)=1
\]
for every pair of settings.
Within this probability space, local correlations are those admitting a
local hidden-variable decomposition~\cite{Bell:1964kc,Bell:1975uz},
\[
\mathcal{L}=
\left\{
P:
P(a,b|x,y)
=
\sum_{\lambda} q(\lambda)
P_A(a|x,\lambda)P_B(b|y,\lambda)
\right\}~,
\]
where $q(\lambda)$ is a probability distribution and $P_A$ and $P_B$ are
local response functions. Measurement independence is assumed, so
$q(\lambda)$ does not depend on $x$ or $y$. Superdeterministic scenarios are
therefore excluded~\cite{Hall:2010zzf,Hossenfelder:2019shy}.

Quantum correlations are those obtainable by local measurements on a
bipartite quantum state~\cite{Tsirelson:1980},
\[
\mathcal{Q}=
\left\{
P:
P(a,b|x,y)
=
\operatorname{Tr}\!\left[
\rho\,
M^{(A)}_{a|x}\otimes M^{(B)}_{b|y}
\right]
\right\}~,
\]
where $\rho$ is a bipartite density operator and
$\{M^{(A)}_{a|x}\}_a$ and $\{M^{(B)}_{b|y}\}_b$ are local
positive-operator-valued measures (POVMs), satisfying
\[
M^{(A)}_{a|x}\geq0~,\qquad
\sum_a M^{(A)}_{a|x}=\mathbb{I}~,
\]
and similarly for the second site.

Imposing only positivity, normalisation, and the no-signalling conditions
\[
\sum_bP(a,b|x,y)
=
\sum_bP(a,b|x,y')~,
\qquad \forall\,a,x,y,y'~,
\]
and
\[
\sum_aP(a,b|x,y)
=
\sum_aP(a,b|x',y)~,
\qquad \forall\,b,x,x',y~,
\]
defines the no-signalling set $\mathcal{NS}$.
The three correlation sets satisfy
\[
\mathcal{L}
\subseteq
\mathcal{Q}
\subseteq
\mathcal{NS}~.
\]
The post-quantum region is the complement of the quantum set within the
no-signalling space,
\[
\mathcal{PQ}
=
\mathcal{NS}
\setminus
\mathcal{Q}~.
\]

The numerical quantities introduced below require a measure $\mu$ on the
no-signalling reference space. Throughout this work, $\mu$ is the Lebesgue
measure in the explicitly specified independent coordinates, so the relevant
fractions are ordinary geometric volume fractions. We evaluate them by uniform
Monte Carlo sampling in these coordinates.

\subsection{Projections}

The complete probability space is generally too high-dimensional for direct
visualisation or efficient experimental analysis. It is therefore useful to
introduce a projection
\[
\pi:
\mathcal{NS}
\longrightarrow
\Pi~,
\]
where $\Pi$ denotes a lower-dimensional representation.
The best-known example is the CHSH projection~\cite{CHSH1969}, which maps a
$(2,2,2,2)$ probability distribution onto a single Bell correlator. Other
projections may retain several independent observables. An example is the
pyramid projection recently constructed for three-setting Bell scenarios with
symmetry-related sites~\cite{Gazdzicki:2026grg}.

Under a projection, the images of the correlation sets become
\[
\pi(\mathcal{L})~,
\qquad
\pi(\mathcal{Q})~,
\qquad
\pi(\mathcal{NS})~,
\]
which satisfy
\[
\pi(\mathcal{L})
\subseteq
\pi(\mathcal{Q})
\subseteq
\pi(\mathcal{NS})~,
\]

Because different points in the complete space may have the same image, a
projection generally reduces the ability to distinguish local, quantum, and
post-quantum correlations.

\section{Popularity, Projection Acceptance and Performance}
\label{sec:popularity}

We now define three quantities that characterise a search for post-quantum
correlations: popularity rate, projection acceptance, and search performance.

\subsection{Popularity rate of post-quantum correlations}
\label{sec:popularity_rate}

The relative abundance of post-quantum correlations depends on the geometry
and dimension of the chosen Bell scenario~\cite{deRosier:2019tud,Lin:2021bne}.
For $\mathcal{PQ}=\mathcal{NS}\setminus\mathcal{Q}$ and a measure $\mu$ on the
reference space $\mathcal{NS}$, we define the \emph{popularity rate} as
\begin{equation}
\label{eq:popularity-rate}
r
=
\frac{\mu(\mathcal{PQ})}
     {\mu(\mathcal{NS})}~.
\end{equation}
Since $\mathcal{NS}$ is the disjoint union
\[
\mathcal{NS}
=
\mathcal{Q}\mathbin{\dot\cup}\mathcal{PQ}~,
\]
Eq.~\eqref{eq:popularity-rate} may also be written as
\begin{equation}
\label{eq:popularity}
r
=
1-
\frac{\mu(\mathcal{Q})}
     {\mu(\mathcal{NS})}~.
\end{equation}
Thus $0\leq r\leq1$. Probabilistically, $r$ is the probability that a point
sampled from $\mathcal{NS}$ according to $\mu$ lies in $\mathcal{PQ}$. Its
value depends on the measure, the number of parties, settings, and outcomes,
and any imposed symmetries. Under the adopted Lebesgue measure, it is simply
the no-signalling volume fraction occupied by post-quantum correlations.

\subsection{Projection acceptance}
\label{sec:acceptance}

Dimensional reduction may map a post-quantum correlation into the projected
quantum region, making it impossible to identify from the retained observables.
For a projection $\pi:\mathcal{NS}\rightarrow\Pi$, we therefore define the
hidden and accepted post-quantum regions as
\[
\mathcal{PQ}^{\mathrm{hid}}(\pi)
=
\{P\in\mathcal{PQ}:\pi(P)\in\pi(\mathcal{Q})\}~,
\qquad
\mathcal{PQ}^{\mathrm{acc}}(\pi)
=
\mathcal{PQ}\setminus\mathcal{PQ}^{\mathrm{hid}}(\pi)~,
\]
Provided $\mu(\mathcal{PQ})>0$, the \emph{projection acceptance} is
\begin{equation}
\label{eq:acceptance}
\mathcal{A}(\pi)
=
\frac{\mu\!\left(\mathcal{PQ}^{\mathrm{acc}}(\pi)\right)}
{\mu(\mathcal{PQ})}
=
1-
\frac{\mu\!\left(\mathcal{PQ}^{\mathrm{hid}}(\pi)\right)}
{\mu(\mathcal{PQ})}~.
\end{equation}
Thus $0\leq\mathcal{A}(\pi)\leq1$. The limiting values $1$ and $0$
correspond, respectively, to a lossless projection and to a projection with no
discriminatory power. The identity projection satisfies $\mathcal{A}(I)=1$.
For Monte Carlo samples, $\mathcal{A}(\pi)$ is estimated by the fraction of
post-quantum points whose images lie outside $\pi(\mathcal{Q})$.

\subsection{Performance of post-quantum searches}
Popularity and acceptance must be considered together. A favourable Bell
scenario may be paired with an uninformative projection, or a highly efficient
projection may act on a scenario in which post-quantum correlations are rare.
We therefore define the \emph{search performance} as
\begin{equation}
\label{eq:performance}
\mathcal{S}(\pi)
=
r\,\mathcal{A}(\pi)~,
\end{equation}
where $r$ is the popularity rate. Using Eqs.~\eqref{eq:popularity-rate} and
\eqref{eq:acceptance}, one obtains
\begin{equation}
\label{eq:performance_volume}
\mathcal{S}(\pi)
=
\frac{
\mu\!\left(
\mathcal{PQ}^{\mathrm{acc}}(\pi)
\right)
}{
\mu(\mathcal{NS})
}~.
\end{equation}
Thus $\mathcal{S}(\pi)$ is both the no-signalling volume fraction occupied by
detectable post-quantum correlations and the probability that a sampled point
is post-quantum and remains identifiable after projection. By construction,
$0\leq\mathcal{S}(\pi)\leq r\leq1$, and
$\mathcal{S}(I)=r$. This decomposition separates the intrinsic suitability of
the Bell scenario from the information retained by the projection and provides
a common figure of merit for comparing search strategies.

\section{Results for symmetry-reduced scenarios}
\label{sec:results}

Here we present results on the detectability of post-quantum correlations in symmetry-reduced 
$(2,2,2,2)$ and $(3,3,2,2)$ scenarios.

\subsection{Exchange-symmetric synchronous correlations}

\subsubsection{Definition and reference spaces}

Throughout this section, we consider binary-outcome Bell scenarios with
settings $x,y\in\{0,\ldots,m-1\}$ and outcomes
$a,b\in\{-1,+1\}$. We use the expectation-value parametrisation

\begin{equation}
    P(a,b|x,y)=\frac{1}{4}
\left(1+aA_x+bB_y+abE_{xy}\right)~,
\label{eq:binary-probability}
\end{equation}
where
\begin{equation}
A_x=\langle a\rangle_x~,\qquad
B_y=\langle b\rangle_y~,\qquad
E_{xy}=\langle ab\rangle_{xy}~.
\end{equation}
We first impose exchange symmetry,
\[
P(a,b|x,y)=P(b,a|y,x)~,
\]
which implies
\[
A_x=B_x~,\qquad E_{xy}=E_{yx}~.
\]

We additionally impose synchrony (perfect same-setting
agreement)~\cite{Russell:2020xho,Dykema_2015}. Whenever the same setting is
selected at the two sites, the outcomes coincide with probability one,

\[
P(a\neq b|x,x)=0~.
\]
For binary outcomes this condition is equivalent to
\[
E_{xx}=1~.
\]
This is an operational condition on the observed probabilities and does not
assume the existence of two identical physical systems. Correlations satisfying
both this condition and exchange symmetry will be called
exchange-symmetric synchronous correlations.
The two requirements are kept explicit because synchrony alone need not imply
exchange symmetry for a general no-signalling correlation.

We denote the corresponding exchange-symmetric synchronous no-signalling
reference space by $\mathcal{NS}_{\mathrm{syn}}^{(m)}$. After imposing
normalisation and positivity, its independent coordinates are the $m$ marginal
means

\[
A_0,\ldots,A_{m-1}~,
\]
and the $m(m-1)/2$ off-diagonal correlators

\[
E_{xy},\qquad 0\leq x<y\leq m-1~.
\]
The dimension of this reference space is consequently

\[
d_{\mathrm{syn}}(m)
=
m+\frac{m(m-1)}{2}
=
\frac{m(m+1)}{2}~.
\]
In particular,

\[
d_{\mathrm{syn}}(2)=3~,\qquad
d_{\mathrm{syn}}(3)=6~.
\]
We denote by $\mathcal{L}_{\mathrm{syn}}^{(m)}$ the subset of
$\mathcal{NS}_{\mathrm{syn}}^{(m)}$ admitting a local hidden-variable model.

For dichotomic quantum correlators, Tsirelson's vector representation gives
real unit vectors
${\bf u}_x$ and ${\bf v}_y$ such that

\[
E_{xy}={\bf u}_x\cdot{\bf v}_y~,
\qquad
\|{\bf u}_x\|=\|{\bf v}_y\|=1~.
\]
In the synchronous sector,

\[
E_{xx}={\bf u}_x\cdot{\bf v}_x=1~.
\]
Since ${\bf u}_x$ and ${\bf v}_x$ are unit vectors, this implies
${\bf u}_x={\bf v}_x\equiv{\bf w}_x$.
Thus, for synchronous quantum correlations, the common-vector representation
follows automatically from the synchrony condition, and
\[
E_{xy}={\bf w}_x\cdot{\bf w}_y~.
\]
The correlator matrix is therefore a Gram matrix. We denote the
corresponding quantum subset by $\mathcal{Q}_{\mathrm{syn}}^{(m)}$. The relevant
nesting is $\mathcal{L}_{\mathrm{syn}}^{(m)}
\subseteq
\mathcal{Q}_{\mathrm{syn}}^{(m)}
\subseteq
\mathcal{NS}_{\mathrm{syn}}^{(m)}$~.
The inclusions need not always be strict and depend on the number of
measurement settings, as illustrated below.

Correlations in the exchange-symmetric synchronous no-signalling reference
space that cannot be realised quantum mechanically form

\[
\mathcal{PQ}_{\mathrm{syn}}^{(m)}
=
\mathcal{NS}_{\mathrm{syn}}^{(m)}
\setminus
\mathcal{Q}_{\mathrm{syn}}^{(m)}~.
\]
Accordingly, their popularity rate is
\begin{equation}
r_{\mathrm{syn}}^{(m)}
=
\frac{
\mu_{d_{\mathrm{syn}}(m)}
\left(\mathcal{PQ}_{\mathrm{syn}}^{(m)}\right)
}{
\mu_{d_{\mathrm{syn}}(m)}
\left(\mathcal{NS}_{\mathrm{syn}}^{(m)}\right)
}~.
\end{equation}

\subsubsection{Degeneracy of the synchronous \texorpdfstring{$(2,2,2,2)$}{(2,2,2,2)} scenario}
\label{sec:degenerate-2222}

For two settings per site, synchronous correlations satisfy
\begin{equation}
 E_{00}=E_{11}=1~,
 \qquad
 E_{01}=E_{10}=X~,
 \qquad
 -1\leq X\leq1~.
 \label{eq:2222-syn-correlators}
\end{equation}
Together with the two marginal means, these relations give the three
independent coordinates
\begin{equation}
 z_{\mathrm{syn}}=(A_0,A_1,X)~,
 \label{eq:2222-syn-coordinates}
\end{equation}
For equal settings, Eq.~\eqref{eq:binary-probability} gives
\begin{equation}
 P(a,b|x,x)=0
 \quad\text{for }a\neq b~,
 \qquad
 P(a,a|x,x)=\frac{1}{2}(1+aA_x)~,
 \label{eq:2222-diagonal-probabilities}
\end{equation}
where the expression for $P(a,a|x,x)$ follows from $A_x=B_x$, $E_{xx}=1$,
and $a^2=1$.
For the off-diagonal setting pairs,
\begin{equation}
 P(a,b|0,1)
 =\frac{1}{4}(1+aA_0+bA_1+abX)~,
 \label{eq:2222-cross-probability}
\end{equation}
and exchange symmetry gives
$P(a,b|1,0)=P(b,a|0,1)$.

The four positivity conditions for
Eq.~\eqref{eq:2222-cross-probability} allow us to interpret the
off-diagonal probabilities as a joint distribution for the predetermined
answers $s_0,s_1=\pm1$:
\begin{equation}
 q(s_0,s_1)
 \equiv P(s_0,s_1|0,1)
 =\frac{1}{4}
 \left(1+s_0A_0+s_1A_1+s_0s_1X\right)
 \geq0~.
 \label{eq:2222-hidden-weights}
\end{equation}
The four non-negative weights $q(s_0,s_1)$ sum to one and therefore define a
probability distribution over the deterministic assignments $(s_0,s_1)$. Draw
$\lambda=(s_0,s_1)$ with probability $q(s_0,s_1)$ and let both sites return $s_x$
when setting $x$ is selected. The resulting conditional distribution is
\begin{equation}
 P(a,b|x,y)
 =
 \sum_{s_0,s_1}
 q(s_0,s_1)
 \delta_{a,s_x}\delta_{b,s_y}~.
 \label{eq:2222-local-decomposition}
\end{equation}
It reproduces both the equal-setting probabilities in
Eq.~\eqref{eq:2222-diagonal-probabilities} and the off-diagonal probabilities in
Eq.~\eqref{eq:2222-cross-probability}. Consequently, every correlation in the
three-dimensional exchange-symmetric synchronous no-signalling reference space
is local. Since every local correlation is quantum, the three sets coincide:
\begin{equation}
 \mathcal{L}_{\mathrm{syn}}^{(2)}
 =
 \mathcal{Q}_{\mathrm{syn}}^{(2)}
 =
 \mathcal{NS}_{\mathrm{syn}}^{(2)}~.
 \label{eq:2222-set-equality}
\end{equation}
Geometrically, this common set is the tetrahedron
\begin{equation}
 \mathcal{NS}_{\mathrm{syn}}^{(2)}
 =
 \operatorname{conv}
 \left\{
 (1,1,1)~,
 (1,-1,-1)~,
 (-1,1,-1)~,
 (-1,-1,1)
 \right\}
 \label{eq:2222-tetrahedron}
\end{equation}
in the coordinates $(A_0,A_1,X)$. Its vertices correspond to the four deterministic
assignments $(s_0,s_1)$.

The post-quantum region within this reference space is therefore empty,
\begin{equation}
 \mathcal{PQ}_{\mathrm{syn}}^{(2)}
 =
 \mathcal{NS}_{\mathrm{syn}}^{(2)}\setminus \mathcal{Q}_{\mathrm{syn}}^{(2)}
 =\varnothing~,
 \label{eq:2222-empty-pq}
\end{equation}
and the corresponding popularity rate vanishes exactly:
\begin{equation}
 r_{\mathrm{syn}}^{(2)}
 =
 \frac{
 \mu_3\!\left(\mathcal{PQ}_{\mathrm{syn}}^{(2)}\right)
 }{
 \mu_3\!\left(\mathcal{NS}_{\mathrm{syn}}^{(2)}\right)
 }
 =0~.
 \label{eq:2222-zero-popularity}
\end{equation}
This result is exact and requires neither a quantum-membership approximation nor
a numerical integration.

The same degeneracy appears under the CHSH projection. The projected coordinate is
\begin{equation}
 C_{\CHSH}
 =E_{00}+E_{01}+E_{10}-E_{11}
 =2X~.
 \label{eq:2222-chsh-coordinate}
\end{equation}
Since $-1\leq X\leq1$~,
\begin{equation}
 \pi_{\CHSH}\!\left(\mathcal{L}_{\mathrm{syn}}^{(2)}\right)
 =
 \pi_{\CHSH}\!\left(\mathcal{Q}_{\mathrm{syn}}^{(2)}\right)
 =
 \pi_{\CHSH}\!\left(\mathcal{NS}_{\mathrm{syn}}^{(2)}\right)
 =[-2,2]~.
 \label{eq:2222-projected-equality}
\end{equation}
The projection therefore has no power to distinguish local, quantum, and
no-signalling correlations within this reference space.

Because $\mathcal{PQ}_{\mathrm{syn}}^{(2)}$ is empty, the projection acceptance is not defined:
both its numerator and denominator vanish. The search performance is nevertheless
well defined and equals
\begin{equation}
 \mathcal{S}_{\mathrm{syn}}^{(2)}(\pi_{\CHSH})=0~.
 \label{eq:2222-zero-performance}
\end{equation}

This degeneracy is specific to the two-setting case. The
single off-diagonal distribution in Eq.~\eqref{eq:2222-cross-probability} already
provides a joint distribution for all predetermined outcomes. No additional
compatibility condition remains. With three or more settings, several pairwise
distributions must be compatible with one global assignment, and positivity of the
individual pairs is no longer sufficient. Strict separations between the local,
quantum, and no-signalling sets can then occur.
\subsubsection{The non-degenerate synchronous \texorpdfstring{$(3,3,2,2)$}{(3,3,2,2)} scenario}
\label{sec:syn-3322}

For three settings per site, the exchange-symmetric synchronous reference space
satisfies
\begin{equation}
 E_{00}=E_{11}=E_{22}=1~,
 \qquad
 E_{xy}=E_{yx}~,
\end{equation}
and the six independent coordinates can be chosen as
\begin{equation}
 z_{\mathrm{syn}}
 =
 (A_0,A_1,A_2,X,Y,Z)~,
 \qquad
 X=E_{01}~,\quad Y=E_{02}~,\quad Z=E_{12}~.
 \label{eq:3322-syn-coordinates}
\end{equation}
Unlike the two-setting case, the three pairwise distributions need not be
compatible with a single global assignment of outcomes. This allows for
non-trivial separations between the local, quantum, and no-signalling sets.

The equal-setting probabilities are fixed by synchrony as in
Eq.~\eqref{eq:2222-diagonal-probabilities}. For each pair of distinct settings
$i<j$, positivity of Eq.~\eqref{eq:binary-probability} requires
\begin{equation}
 1+s_iA_i+s_jA_j+s_is_jE_{ij}\geq0~,
 \qquad
 s_i,s_j\in\{-1,+1\}~,\qquad i<j ~.
 \label{eq:3322-pair-positivity}
\end{equation}
These conditions guarantee positivity of each observed pairwise distribution,
but not the existence of a joint distribution for all three settings.

A point belongs to the local subset $\mathcal{L}_{\mathrm{syn}}^{(3)}$ if there exists a
triple moment $T\in[-1,1]$ such that
\begin{equation}
 q(s_0,s_1,s_2)
 =
 \frac18
 \left[
 1+\sum_{i=0}^{2}s_iA_i
 +\sum_{0\leq i<j\leq2}s_is_jE_{ij}
 +s_0s_1s_2T
 \right]
 \geq0
 \label{eq:3322-joint-distribution}
\end{equation}
for every $(s_0,s_1,s_2)\in\{-1,+1\}^3$. The distribution $q$ then provides
a local hidden-variable model in which both sites use the same deterministic
assignment $(s_0,s_1,s_2)$.

For the quantum subset $\mathcal{Q}_{\mathrm{syn}}^{(3)}$, synchrony implies
the common-vector representation derived above. Introducing a unit reference
vector ${\bf w}_{\ast}$ for the marginal means, one may write
\begin{equation}
 A_i={\bf w}_{\ast}\cdot{\bf w}_i~,
 \qquad
 E_{ij}={\bf w}_i\cdot{\bf w}_j~.
 \label{eq:3322-augmented-vectors}
\end{equation}
Thus the synchronous quantum correlations admit
positive semidefiniteness of the
augmented Gram matrix
\begin{equation}
 \Gamma(A_0,A_1,A_2,X,Y,Z)
 =
 \begin{pmatrix}
 1   & A_0 & A_1 & A_2\\
 A_0 & 1   & X   & Y\\
 A_1 & X   & 1   & Z\\
 A_2 & Y   & Z   & 1
 \end{pmatrix}
 \succeq0~.
 \label{eq:3322-augmented-gram}
\end{equation}
But the converse does not follow: a positive Gram matrix with the prescribed
marginal means need not have a synchronous quantum realisation.
Thus, using condition~\ref{eq:3322-augmented-gram}, one estimates an upper bound on quantum correlations ($\mathcal{Q}_{\mathrm{syn}}^{(3)} \subsetneq \tilde{\mathcal{Q}}_{\mathrm{syn}}^{(3)}$) and a lower bound on post-quantum correlations (the Gram-matrix certified post-quantum correlations,
$\tilde{\mathcal{PQ}}_{\mathrm{syn}}^{(3)})$.

The three sets of correlations satisfy
\begin{equation}
 \mathcal{L}_{\mathrm{syn}}^{(3)}
 \subsetneq
 \mathcal{Q}_{\mathrm{syn}}^{(3)}
 \subsetneq
 \mathcal{NS}_{\mathrm{syn}}^{(3)}~.
 \label{eq:3322-strict-nesting}
\end{equation}
The origin of the difference from the $(2,2,2,2)$ case is already visible from
the example
\begin{equation}
 X=Y=Z=-1~.
\end{equation}
Each pair separately describes valid perfect anti-correlation, but three binary
variables cannot be pairwise opposite. Equivalently, the corresponding
three-vector Gram matrix is not positive semidefinite.
The post-quantum region is
\begin{equation}
 \mathcal{PQ}_{\mathrm{syn}}^{(3)}
 =
 \mathcal{NS}_{\mathrm{syn}}^{(3)}\setminus \mathcal{Q}_{\mathrm{syn}}^{(3)}~.
 \label{eq:3322-post-quantum-set}
\end{equation}
We estimate its six-dimensional volume by rejection sampling. Candidate points
are generated uniformly in $[-1,1]^6$ and retained when all inequalities in
Eq.~\eqref{eq:3322-pair-positivity} are satisfied. Locality is tested through
Eq.~\eqref{eq:3322-joint-distribution}, while quantum membership is tested
through Eq.~\eqref{eq:3322-augmented-gram}.

Using $2.0\times10^7$ candidates and the fixed pseudorandom seed $20260807$, we
obtain the counts in Table~\ref{tab:3322-set-counts}.

\begin{table}[tbp]
 \centering
 \caption{Monte Carlo counts in the six-dimensional exchange-symmetric synchronous reference space.}
 \label{tab:3322-set-counts}
 \vspace{0.5cm}
 \begin{tabular}{@{}lr@{}}
 \toprule
 Set or subset & Number of retained points\\
 \midrule
 $\mathcal{NS}_{\mathrm{syn}}^{(3)}$ & $1\,333\,382$\\
 $\mathcal{L}_{\mathrm{syn}}^{(3)}$  & $889\,036$\\
 $\tilde{\mathcal{Q}}_{\mathrm{syn}}^{(3)}$  & $1\,205\,116$\\
 $\tilde{\mathcal{PQ}}_{\mathrm{syn}}^{(3)}$ & $128\,266$\\
 \bottomrule
 \end{tabular}
\end{table}
The local and quantum volume fractions are
\begin{equation}
 \frac{\mu_6(\mathcal{L}_{\mathrm{syn}}^{(3)})}
 {\mu_6(\mathcal{NS}_{\mathrm{syn}}^{(3)})}
 =0.66675\pm0.00080~,
 \label{eq:3322-local-fraction}
\end{equation}
and
\begin{equation}
 \frac{\mu_6(\tilde{\mathcal{Q}}_{\mathrm{syn}}^{(3)})}
 {\mu_6(\mathcal{NS}_{\mathrm{syn}}^{(3)})}
 =0.90380\pm0.00050~,
 \label{eq:3322-quantum-fraction}
\end{equation}
respectively. The resulting popularity rate (certified lower bound) is
\begin{equation}
 \widehat r_{\mathrm{syn}}^{(3)}
 =
 \frac{128\,266}{1\,333\,382}
 =0.09620\pm0.00050~,
 \label{eq:3322-popularity-result}
\end{equation}
where the quoted uncertainty is a 95\% Monte Carlo confidence interval. Thus,
approximately $9.62\%$ of the six-dimensional reference space is post-quantum.

We next consider the pyramid projection, which retains only the three
off-diagonal correlators \cite{Gazdzicki:2026grg},
\begin{equation}
 \pi_{\Pyr}:
 (A_0,A_1,A_2,X,Y,Z)
 \longmapsto
 (X,Y,Z)~,
 \label{eq:3322-pyramid-projection}
\end{equation}
The projected local set is the tetrahedron
\begin{equation}
 \pi_{\Pyr}(\mathcal{L}_{\mathrm{syn}}^{(3)})
 =
 \operatorname{conv}
 \left\{
 (1,1,1)~,
 (1,-1,-1)~,
 (-1,1,-1)~,
 (-1,-1,1)
 \right\}~,
 \label{eq:3322-projected-local-set}
\end{equation}
whereas the projected quantum set is the elliptope
\begin{equation}
 \pi_{\Pyr}(\mathcal{Q}_{\mathrm{syn}}^{(3)})
 =
 \left\{
 (X,Y,Z)\in[-1,1]^3:
 1+2XYZ-X^2-Y^2-Z^2\geq0
 \right\}~,
 \label{eq:3322-projected-quantum-set}
\end{equation}
and
\begin{equation}
 \pi_{\Pyr}(\mathcal{NS}_{\mathrm{syn}}^{(3)})=[-1,1]^3~.
 \label{eq:3322-projected-ns-set}
\end{equation}

Defining
\begin{equation}
 \Delta_{\Pyr}(X,Y,Z)
 =
 \det
 \begin{pmatrix}
  1&X&Y\\
  X&1&Z\\
  Y&Z&1
 \end{pmatrix}
 =
 1+2XYZ-X^2-Y^2-Z^2~,
 \label{eq:3322-pyramid-determinant}
\end{equation}
a post-quantum point is identified by the pyramid projection whenever
\begin{equation}
 \Delta_{\Pyr}(X,Y,Z)<0~.
 \label{eq:3322-pyramid-acceptance-condition}
\end{equation}
For the $128\,266$ post-quantum points in the Monte Carlo sample, we find
\begin{equation}
 N_{\mathrm{acc}}=112\,551~,
 \qquad
 N_{\mathrm{hid}}=15\,715~.
 \label{eq:3322-acceptance-counts}
\end{equation}
The corresponding upper-bound estimate for the true projection acceptance) is
\begin{equation}
 \widehat{\mathcal{A}}_{\mathrm{syn}}^{(3)}(\pi_{\Pyr})
 =
 \frac{112\,551}{128\,266}
 =0.87748\pm0.00179~,
 \label{eq:3322-acceptance-result}
\end{equation}
while the search performance estimate is
\begin{equation}
 \widehat{\mathcal{S}}_{\mathrm{syn}}^{(3)}(\pi_{\Pyr})
 =
 \widehat r_{\mathrm{syn}}^{(3)}
 \widehat{\mathcal{A}}_{\mathrm{syn}}^{(3)}(\pi_{\Pyr})
 =
 \frac{112\,551}{1\,333\,382}
 =0.08441\pm0.00047~.
 \label{eq:3322-performance-result}
\end{equation}
Thus, approximately $87.75\%$ of the post-quantum correlations remain
identifiable after projection, corresponding to approximately $8.44\%$ of the
complete reference space.

\begin{figure}[htbp]
 \centering
 \includegraphics[width=\linewidth]{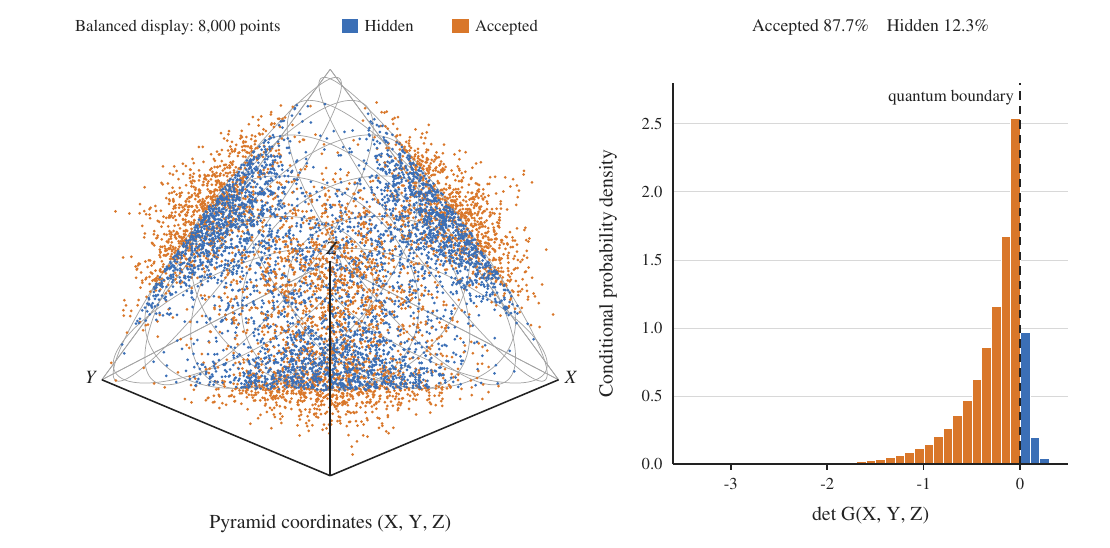}
 \caption{Distribution of the Gram-matrix certified post-quantum exchange-symmetric synchronous
 correlations in the pyramid representation. \textit{Left:} a balanced
 visualisation subsample of $8\,000$ points,
 with $4\,000$ hidden and $4\,000$ accepted points. The grey wireframe denotes
 the elliptope boundary $\Delta_{\Pyr}=0$. \textit{Right:} distribution of
 $\Delta_{\Pyr}$ for all $128\,266$ post-quantum points. The accepted and hidden
 fractions are $0.87748$ and $0.12252$, respectively.}
 \label{fig:3322-pyramid-distribution}
\end{figure}

Figure~\ref{fig:3322-pyramid-distribution} illustrates the distribution of these
post-quantum points in the pyramid representation. The three-dimensional panel
uses a balanced subsample for visual clarity, whereas the determinant histogram
uses the complete post-quantum sample.

\begin{figure}[htbp]
 \centering
 \includegraphics[width=\linewidth]{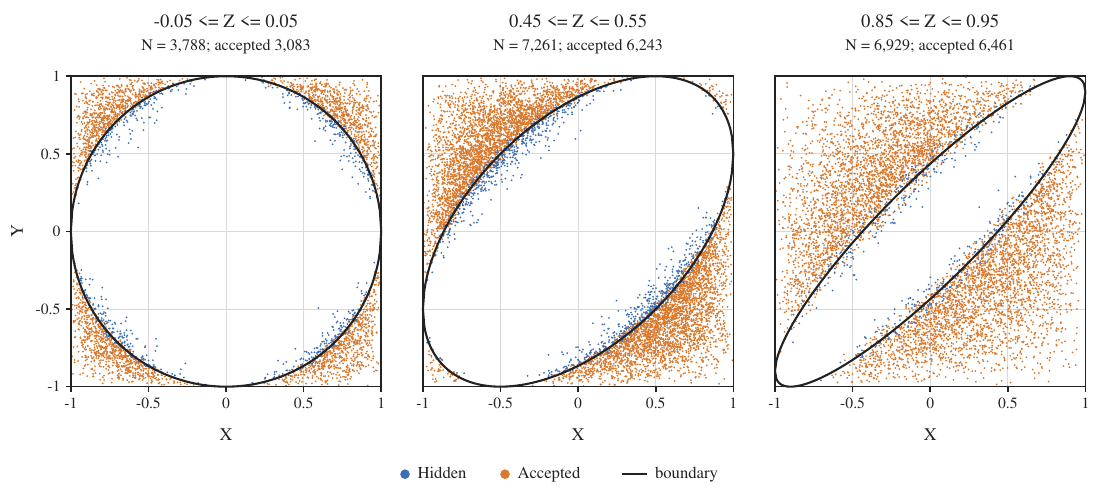}
 \caption{Certified post-quantum exchange-symmetric synchronous correlations in three
 $X$--$Y$ slices of the pyramid representation, centred at $Z=0$, $Z=0.5$, and
 $Z=0.9$. Blue points
 lie inside the projected quantum elliptope and are hidden by the projection.
 Orange points lie outside it and are identified as post-quantum. The solid
 curves denote the corresponding central elliptope boundaries.}
 \label{fig:3322-pyramid-slices}
\end{figure}

For additional visualisation, Fig.~\ref{fig:3322-pyramid-slices} shows three
$X$--$Y$ slices centred at $Z=0$, $Z=0.5$, and $Z=0.9$. At fixed $Z=z_0$, the
elliptope boundary is
\begin{equation}
 Y
 =
 Xz_0
 \pm
 \sqrt{(1-X^2)(1-z_0^2)}~.
 \label{eq:3322-pyramid-slice-contour}
\end{equation}
The figures show that some post-quantum points are hidden inside the projected
quantum elliptope, although the pyramid representation retains most of them.
\subsection{Exchange-symmetric correlations without synchrony}
\label{sec:exchange-symmetric-examples}

We now retain exchange symmetry,
\begin{equation}
 A_x=B_x~,
 \qquad
 E_{xy}=E_{yx}~,
 \label{eq:ex-moment-symmetry}
\end{equation}
but drop the synchrony condition imposed in the preceding subsection.
Consequently, the diagonal correlators $E_{xx}$ are no longer fixed to one and
remain independent coordinates. We denote the corresponding no-signalling,
local, and quantum sets by $\NSex^{(m)}$, $\Lex^{(m)}$, and $\Qex^{(m)}$,
respectively, with
\begin{equation}
 \Lex^{(m)}
 \subseteq
 \Qex^{(m)}
 \subseteq
 \NSex^{(m)}~,
 \qquad
 \PQex^{(m)}
 =
 \NSex^{(m)}\setminus\Qex^{(m)}~.
 \label{eq:ex-set-nesting}
\end{equation}
The independent coordinates are the $m$ marginal means and the $m(m+1)/2$
independent correlators, giving
\begin{equation}
 d_{\mathrm{ex}}(m)
 =
 \frac{m(m+3)}{2}~.
 \label{eq:ex-reference-dimension}
\end{equation}
Thus $d_{\mathrm{ex}}(2)=5$ and $d_{\mathrm{ex}}(3)=9$.

\subsubsection{The exchange-symmetric \texorpdfstring{$(2,2,2,2)$}{(2,2,2,2)} scenario}
\label{sec:exchange-symmetric-2222}

For two settings per site, the five independent coordinates are
\begin{equation}
 z_{\mathrm{ex}}
 =
 (A_0,A_1,E_{00},E_{01},E_{11})~,
 \qquad
 E_{10}=E_{01}~.
 \label{eq:ex-2222-coordinates}
\end{equation}
The probabilities in Eq.~\eqref{eq:binary-probability} must be non-negative.
In contrast to the synchronous case, $E_{00}$ and $E_{11}$ are free.

For the numerical estimate below, we certify a subset of post-quantum
correlations using an outer approximation to $\Qex^{(2)}$. For
non-deterministic marginals, one defines
\begin{equation}
 D_{xy}
 =
 \frac{E_{xy}-A_xA_y}
 {\sqrt{(1-A_x^2)(1-A_y^2)}}~.
 \label{eq:ex-2222-normalised-correlators}
\end{equation}
Cases in which a denominator vanishes occur only on the boundary of the
five-dimensional reference space and do not affect the volume estimates.
A necessary condition for quantum realisability is~\cite{Navascues:2007zcm}
\begin{equation}
 \left|
 \sum_{x,y=0}^{1}
 s_{xy}\arcsin D_{xy}
 \right|
 \leq\pi
 \label{eq:ex-2222-arcsine-condition}
\end{equation}
for every choice $s_{xy}\in\{-1,+1\}$ satisfying
\begin{equation}
 \prod_{x,y=0}^{1}s_{xy}=-1~,
 \label{eq:ex-2222-sign-condition}
\end{equation}
together with $|D_{xy}|\leq1$. Equation~\eqref{eq:ex-2222-arcsine-condition}
therefore represents a family of inequalities. Because exchange symmetry gives
$D_{10}=D_{01}$, only three of them are distinct. A point violating any one of
these inequalities, or one of the bounds $|D_{xy}|\leq1$, fails this necessary
quantum condition. It is therefore certainly post-quantum. We denote the
corresponding outer approximation to the quantum set by $\Qoneex^{(2)}$ and the
certified post-quantum subset by
\begin{equation}
 \PQcert^{(2)}
 =
 \NSex^{(2)}\setminus\Qoneex^{(2)}
 \subseteq
 \PQex^{(2)}~.
 \label{eq:ex-2222-certified-subset}
\end{equation}

We sample uniformly in the five independent coordinates of
Eq.~\eqref{eq:ex-2222-coordinates}. From $2.0\times10^7$ points generated in
$[-1,1]^5$, using the fixed pseudorandom seed $20260807$, we retain
\begin{equation}
 N_{\mathrm{NS}}=2\,667\,285~,
 \qquad
 N_{\mathcal{PQ}}^{(1)}=46\,484~.
 \label{eq:ex-2222-sample-counts}
\end{equation}
The resulting certified popularity is
\begin{equation}
 \widehat r_1^{(2)}
 =
 \frac{N_{\mathcal{PQ}}^{(1)}}{N_{\mathrm{NS}}}
 =
 0.01743\pm0.00016~.
 \label{eq:ex-2222-certified-popularity}
\end{equation}
Thus, approximately $1.74\%$ of the exchange-symmetric no-signalling reference
space is certified as post-quantum by this test. Since the test is based on an
outer approximation to the quantum set,
$r_{\mathrm{ex}}^{(2)}
 \geq
 \widehat r_1^{(2)}$.
The quoted $95\%$ confidence interval describes Monte Carlo sampling uncertainty
only, not the systematic difference between the outer approximation and the
complete quantum set.

\begin{figure}[htbp]
 \centering
 \includegraphics[width=0.94\linewidth]{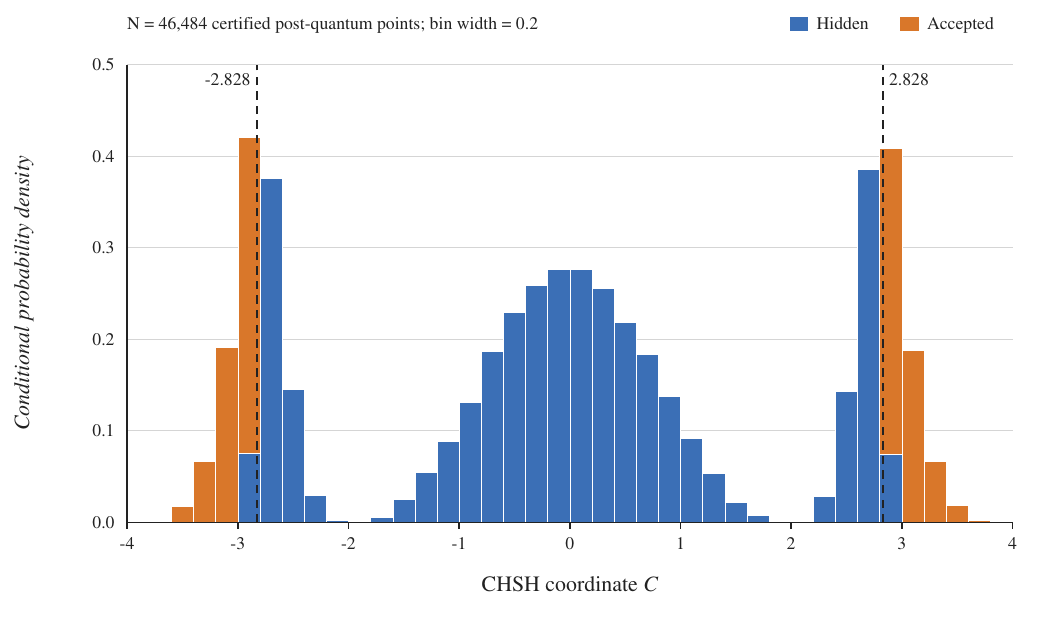}
 \caption{Distribution of
 $C_{\CHSH}=E_{00}+2E_{01}-E_{11}$ for the $46\,484$
 exchange-symmetric correlations certified as post-quantum. The dashed lines
 mark the Tsirelson boundaries $C_{\CHSH}=\pm2\sqrt2$. Correlations outside
 these boundaries are identified by the CHSH projection, whereas those inside
 the projected quantum interval are hidden.}
 \label{fig:ex-2222-chsh-distribution}
\end{figure}

We now apply the CHSH projection,
\begin{equation}
 C_{\CHSH}
 =
 E_{00}+E_{01}+E_{10}-E_{11}
 =
 E_{00}+2E_{01}-E_{11}~.
 \label{eq:ex-2222-chsh-coordinate}
\end{equation}
The projected local, quantum, and no-signalling sets are
\begin{equation}
 \pi_{\CHSH}\!\left(\Lex^{(2)}\right)=[-2,2]~,
 \qquad
 \pi_{\CHSH}\!\left(\Qex^{(2)}\right)
 =[-2\sqrt2,2\sqrt2]~,
 \qquad
 \pi_{\CHSH}\!\left(\NSex^{(2)}\right)=[-4,4]~.
 \label{eq:ex-2222-projected-sets}
\end{equation}
Thus, exchange symmetry does not reduce the usual Tsirelson bound.
For example, the exchange-symmetric quantum correlator matrix
\begin{equation}
 \begin{pmatrix}
 E_{00}&E_{01}\\
 E_{10}&E_{11}
 \end{pmatrix}
 =
 \frac{1}{\sqrt2}
 \begin{pmatrix}
 1&1\\
 1&-1
 \end{pmatrix}
 \label{eq:ex-2222-tsirelson-point}
\end{equation}
attains $C_{\CHSH}=2\sqrt2$.
A certified post-quantum point is therefore identified by the CHSH projection
when
\begin{equation}
 |C_{\CHSH}|>2\sqrt2~.
 \label{eq:ex-2222-acceptance-condition}
\end{equation}
Among the $46\,484$ certified post-quantum points, $11\,462$ satisfy this
condition, while $35\,022$ are projected inside the quantum interval. Hence
\begin{equation}
 \widehat{\mathcal{A}}_1(\pi_{\CHSH})
 =
 \frac{11\,462}{46\,484}
 =
 0.2466\pm0.0039~,
 \label{eq:ex-2222-certified-acceptance}
\end{equation}
and
\begin{equation}
 \widehat{\mathcal{S}}_1(\pi_{\CHSH})
 =
 \frac{11\,462}{2\,667\,285}
 =
 0.00430\pm0.00008~.
 \label{eq:ex-2222-certified-performance}
\end{equation}
Thus CHSH identifies approximately $24.7\%$ of the correlations certified as
post-quantum by the present test, corresponding to approximately $0.43\%$ of
the complete exchange-symmetric reference space. The conditional
acceptance, however, need not be a lower bound on the exact acceptance, since
the post-quantum correlations not certified by the present test may have a
different CHSH distribution.
We collect the numerical results in Table~\ref{tab:ex-2222-summary}.
\begin{table}[htbp]
 \centering
 \caption{Monte Carlo results for the exchange-symmetric $(2,2,2,2)$ scenario.
 The uncertainties are $95\%$ Monte Carlo confidence intervals and refer to the
 certified sample.}
 \label{tab:ex-2222-summary}
 \vspace{0.5cm}
 \begin{tabular}{@{}lr@{}}
 \toprule
  Quantity & Value \\
  \midrule
  Retained no-signalling points, $N_{\mathrm{NS}}$ & $2\,667\,285$ \\
  Certified post-quantum points, $N_{\mathcal{PQ}}^{(1)}$ & $46\,484$ \\
  Certified popularity, $\widehat r_1^{(2)}$ & $0.01743\pm0.00016$ \\
  Accepted certified points & $11\,462$ \\
  Hidden certified points & $35\,022$ \\
  Certified acceptance, $\widehat{\mathcal{A}}_1$ & $0.2466\pm0.0039$ \\
  Certified performance, $\widehat{\mathcal{S}}_1$ & $0.00430\pm0.00008$ \\
  \bottomrule
 \end{tabular}
\end{table}
Figure~\ref{fig:ex-2222-chsh-distribution} shows the distribution of the
certified post-quantum sample along the CHSH coordinate. The two tails outside
the Tsirelson interval contain the accepted correlations, while the correlations
inside the interval are hidden by the projection.

The three-dimensional synchronous space studied above has zero volume within
the present five-dimensional reference space. This does not imply unit
post-quantum popularity: its complement contains many ordinary quantum
correlations. Popularity must be evaluated by comparing $\Qex^{(2)}$ and
$\NSex^{(2)}$ within the same reference space.

\subsubsection{The exchange-symmetric \texorpdfstring{$(3,3,2,2)$}{(3,3,2,2)} scenario}
\label{sec:exchange-symmetric-3322}

For three settings per site, exchange symmetry leaves nine
independent coordinates,
\begin{equation}
 z_{\mathrm{ex}}
 =
 (A_0,A_1,A_2,E_{00},E_{11},E_{22},E_{01},E_{02},E_{12})~,
 \label{eq:ex-3322-coordinates}
\end{equation}
The exchange-symmetric no-signalling set $\NSex^{(3)}$ is determined by
positivity of Eq.~\eqref{eq:binary-probability} for all outcomes and settings.
Again, the diagonal correlators
$E_{00},E_{11},E_{22}$ remain independent.

To obtain a certified lower bound on the post-quantum popularity, we apply the
same two-setting quantum test used above separately to the three setting pairs
$\{0,1\}$, $\{0,2\}$, and $\{1,2\}$. For each pair $i<j$, the normalised
correlators defined in Eq.~\eqref{eq:ex-2222-normalised-correlators} must satisfy
Eq.~\eqref{eq:ex-2222-arcsine-condition} and $|D_{xy}|\leq1$. If any
two-setting restriction violates one of these necessary conditions, the
complete three-setting correlation is certainly post-quantum.

Let $\Qpairex^{(3)}$ denote the set of points passing all three pairwise tests.
Every quantum correlation must pass each restriction, so
$\Qex^{(3)}\subseteq\Qpairex^{(3)}$ and therefore
\begin{equation}
 \PQpair^{(3)}
 =
 \NSex^{(3)}\setminus\Qpairex^{(3)}
 \subseteq
 \PQex^{(3)}~.
 \label{eq:ex-3322-certified-subset}
\end{equation}
Sampling uniformly in the nine independent coordinates, we generate
$2.0\times10^7$ candidate points with the same fixed pseudorandom seed and
obtain
\begin{equation}
 N_{\mathrm{NS}}=476\,495~,
 \qquad
 N_{\mathcal{PQ}}^{(\mathrm{pair})}=28\,801~.
 \label{eq:ex-3322-sample-counts}
\end{equation}
The pairwise-certified popularity is consequently
\begin{equation}
 \widehat r_{\mathrm{pair}}^{(3)}
 =
 \frac{28\,801}{476\,495}
 =
 0.06044\pm0.00068~,
 \label{eq:ex-3322-certified-popularity}
\end{equation}
and provides the lower bound
$r_{\mathrm{ex}}^{(3)}
 \geq
 \widehat r_{\mathrm{pair}}^{(3)}$.
Thus at least approximately $6.04\%$ of the nine-dimensional
exchange-symmetric no-signalling space is post-quantum. As above, the quoted
confidence interval refers only to Monte Carlo sampling.

We finally consider the pyramid projection,
\begin{equation}
 \pi_{\Pyr}:
 z_{\mathrm{ex}}
 \longmapsto
 (X,Y,Z)~,
 \qquad
 X=E_{01}~,\quad
 Y=E_{02}~,\quad
 Z=E_{12}~.
 \label{eq:ex-3322-pyramid-map}
\end{equation}
Positivity implies $|X|,|Y|,|Z|\leq1$, and every point of this cube is realised
by an exchange-symmetric no-signalling correlation. Hence
\begin{equation}
 \pi_{\Pyr}\!\left(\NSex^{(3)}\right)=[-1,1]^3~.
 \label{eq:ex-3322-projected-ns-set}
\end{equation}

Remarkably, the complete cube is already the projection of the local set. To
see this, consider an arbitrary vertex
$(X,Y,Z)\in\{-1,+1\}^3$,
and define
\begin{equation}
 \eta =XYZ~,
 \qquad
 \alpha_0=1~,
 \qquad
 \alpha_1=\frac{X}{\eta}~,
 \qquad
 \alpha_2=\frac{Y}{\eta}~,
 \qquad
 \beta_i=\eta\alpha_i~.
 \label{eq:ex-3322-local-signs}
\end{equation}
Let $\lambda\in\{-1,+1\}$ be uniformly distributed and choose the local
responses
\begin{equation}
 a=\alpha_x\lambda~,
 \qquad
 b=\beta_y\lambda~.
 \label{eq:ex-3322-local-responses}
\end{equation}
They give
\begin{equation}
 A_x=B_x=0~,
 \qquad
 E_{xy}
 =
 \alpha_x\beta_y
 =
 r\alpha_x\alpha_y
 =
 E_{yx}~,
 \label{eq:ex-3322-local-correlators}
\end{equation}
with projected coordinates precisely $(X,Y,Z)$. Thus every cube vertex has an
exchange-symmetric local realisation. By convexity,
\begin{equation}
 \pi_{\Pyr}\!\left(\Lex^{(3)}\right)
 =
 \pi_{\Pyr}\!\left(\Qex^{(3)}\right)
 =
 \pi_{\Pyr}\!\left(\NSex^{(3)}\right)
 =
 [-1,1]^3~.
 \label{eq:ex-3322-projected-set-equality}
\end{equation}

The pyramid projection therefore has no discriminatory power in the
exchange-symmetric model. Since every projected point already belongs to the
projected local, and hence quantum, set, no post-quantum correlation can be
identified from $(X,Y,Z)$ alone. Consequently,
\begin{equation}
 \mathcal{A}_{\mathrm{ex}}^{(3)}(\pi_{\Pyr})=0~,
 \qquad
 \mathcal{S}_{\mathrm{ex}}^{(3)}(\pi_{\Pyr})=0~.
 \label{eq:ex-3322-zero-acceptance-performance}
\end{equation}
The non-zero certified popularity found above ensures that the first of these
quantities is well defined: post-quantum correlations exist, but all of them
are hidden by the projection.
Table~\ref{tab:ex-3322-summary} summarises the results.
\begin{table}[htbp]
 \centering
 \caption{Results for the exchange-symmetric $(3,3,2,2)$ scenario and the
 pyramid projection. The popularity is a pairwise-certified lower bound with a
 $95\%$ Monte Carlo confidence interval, whereas the vanishing projection
 acceptance and performance are exact.}
 \label{tab:ex-3322-summary}
 \vspace{0.5cm}
 \begin{tabular}{@{}lr@{}}
  \toprule
  Quantity & Value \\
  \midrule
  Reference-space dimension & $9$ \\
  Retained no-signalling points, $N_{\mathrm{NS}}$ & $476\,495$ \\
  Pairwise-certified post-quantum points & $28\,801$ \\
  Certified popularity, $\widehat r_{\mathrm{pair}}^{(3)}$
      & $0.06044\pm0.00068$ \\
  Pyramid acceptance, $\mathcal{A}_{\mathrm{ex}}^{(3)}$ & $0$ \\
  Pyramid performance, $\mathcal{S}_{\mathrm{ex}}^{(3)}$ & $0$ \\
  \bottomrule
 \end{tabular}
\end{table}

This result also clarifies the relation to the synchronous case. There,
synchrony leads in quantum theory to the common-vector representation, and the
pyramid projection of the quantum set is the elliptope
$1+2XYZ-X^2-Y^2-Z^2\geq0$.
With exchange symmetry alone, this restriction no longer applies. Points
outside the elliptope but inside the cube need not be post-quantum and can even
be local. Thus the same pyramid representation that retains substantial
discriminatory power in the synchronous sector loses all such power when only
exchange symmetry is imposed.

\section{Summary and conclusions}
\label{sec:conclusions}

We introduced three quantities for characterising searches for post-quantum
correlations: popularity rate, projection acceptance, and their product, search
performance. Popularity measures the abundance of post-quantum correlations
before projection, whereas acceptance measures how much of this information
survives in the reduced representation. We use these concepts to quantify and compare the detectability of post-quantum correlation in several important symmetry-reduced cases.

Needless to say that popularity and acceptance must be treated
consistently. In particular, the reference space and the quantum set used to
define post-quantum correlations must belong to the same physical model.
This distinction is especially clear when comparing exchange-symmetric
synchronous correlations with correlations satisfying only observable exchange
symmetry. The synchrony condition is also observable and, for binary outcomes,
equivalent to $E_{xx}=1$. In quantum theory it implies the common-vector
representation. 

For two settings, the local, quantum, and no-signalling
synchronous sets coincide, so the post-quantum popularity and
the CHSH performance both vanish. For three settings, the situation becomes
non-trivial. The pyramid representation separates the local tetrahedron, the
quantum elliptope, and the no-signalling cube. In this case, using the Gram-matrix certified correlations, we find a
post-quantum popularity of about $9.6\%$, while the pyramid projection retains
about $87.7\%$ of these correlations, corresponding to an overall performance of
about $8.4\%$.

The picture changes substantially when the synchrony condition is removed, and
only exchange symmetry is retained. In the $(2,2,2,2)$ scenario, the
quantum CHSH image remains the usual Tsirelson interval. A first-level outer
test certifies about $1.7\%$ of the exchange-symmetric no-signalling space as
post-quantum, but CHSH identifies only about one quarter of this certified
subset. In the $(3,3,2,2)$ case, pairwise application of the same test certifies
a non-zero post-quantum fraction of about $6.0\%$. Yet the pyramid projection loses all
discriminatory power. Its projected local, quantum, and no-signalling sets all
coincide with the cube.

The exchange-symmetric results based on outer quantum tests are conservative.
The quoted Monte Carlo uncertainties describe sampling errors, while the
remaining systematic uncertainty is associated with the approximation to the
full quantum set. Higher levels of the NPA hierarchy may refine these estimates.
More generally, the present framework can be used to compare alternative
projections.

In future, it will be interesting to apply this framework to
realistic physical cases. Natural cases are provided by collider-produced fermion pairs~\cite{Barr:2024djo,ATLAS:2023fsd} or positronium decays into
photons~\cite{Moskal:2024fpv,Kumar:2023xpx}, where experimentally accessible
correlations can be confronted directly with the geometric structures
discussed here.

\vspace{0.5cm}
\textbf{Acknowledgments}
This work was supported by the Polish Minister of Science under the ‘Regional Excellence Initiative’ program (project RID/SP/0015/2024/01). 

\vspace{0.5cm}
\textbf{AI disclosure:} ChatGPT (OpenAI) assisted in discussing and refining the authors' ideas, revising the text and language, and preparing figures. The scientific ideas and conclusions are those of the authors.

\bibliographystyle{elsarticle-num}
\bibliography{references}

\end{document}